\documentclass{openjournal}
\usepackage{amssymb,amsmath}
\def\app#1#2{%
	\mathrel{%
		\setbox0=\hbox{$#1\approx$}%
		\setbox2=\hbox{%
			\rlap{\hbox{$#1\propto$}}%
			\lower1.1\ht0\box0%
			}%
			\raise0.25\ht2\box2%
			}%
			}

\usepackage{revsymb,graphicx,amsmath,amssymb}
\begin{document}
\title{Triaxial Magnetars as Sources of Fast Radio Bursts}
\shorttitle{Triaxial Magnetars as Sources of FRB}
\shortauthors{Katz}
\author{J. I. Katz\altaffilmark{}}
\affil{Department of Physics and McDonnell Center for the Space Sciences,
Washington University, St. Louis, Mo. 63130}
\email{katz@wuphys.wustl.edu}
\begin{abstract}
Some of the mysterious temporal properties of Fast Radio Bursts (FRB) may be
explained if they are emitted by dynamically triaxial magnetars.  If the
bursts are narrowly collimated along open field lines, then observed
repeating FRB are those few whose rotation axis, open field lines and
infrequent radiation {(analogous to pulsar giant pulses)} point nearly
to the observer.  In apparently non-repeating FRB {these are misaligned
and} the directions of the open field lines and infrequent radiation wander
across the sky {as they rotate}, reducing their observed duty factors by
several orders of magnitude.  In repeaters a triaxial moment tensor moves
the radiation pattern into or out of the line of sight {on long
(precessional) time scales}, explaining periods of greater or lesser (or
absent) activity.  The dynamics of triaxial bodies may thwart the coherent
integration of gravitational signals from rotating neutron stars.
\end{abstract}
\newpage
\section{Introduction}
Many properties of FRB are mysterious, not least why they emit high
brightness radio bursts, a mystery they share with radio pulsars.  Here I
address three other mysteries within the popular and plausible magnetar model
\citep{PP07,PP13,T13,L14,PC15,K16,S16} of FRB:
\begin{enumerate}
\item If the magnetar model of FRB is correct, why have their rotational
periods not been found, unlike those of known magnetar Soft Gamma Repeaters
(SGR) and Anomalous X-ray Pulsars (AXP, which are SGR between outbursts),
whose 2--12 s rotational periods appear as large amplitude modulation of
their emissions?
\item Why does the activity of repeating FRB vary by large factors between
episodes of intense activity and episodes of apparent inactivity?
\item Why do the duty factors of repeating FRB and apparently non-repeating
FRB differ by several orders of magnitude? 
\end{enumerate}

In this paper I attribute these phenomena to a triaxial moment of inertia
tensor of a magnetically and possibly elastically (by stress frozen-in to
the solid component) distorted neutron star.  These differences may be
enhanced if magnetar-FRB have larger magnetic fields than the known Galactic
magnetar-SGR/AXP, implying larger asymmetry of magnetar-FRB moments of
inertia.

{Precessing triaxial neutron stars have been proposed to explain the
slow periodic (16 d and 160 d) activity modulation of FRB 180916B and FRB
121102A \citep{ZL20,C22,W22a}, and these authors presented detailed analyses
of their complex dynamics.  I draw on these analyses here but make no
assumptions about these periodic modulations, for which other explanations
have been proposed
\citep{BWM20,C20,LBB20,K20,S20,Y20,C21a,L21,K21,S21,K22b,W22b,F24,K24a,K24b}.
\citet{C22,W22a}
suggested that torque variation over times of several days and phase jitter
may explain the failure to observe rotational periods of FRB.  However,
\citet{K22a} argued that the required magnitude of phase jitter may be
implausibly large and \citet{K25b} found no rotational period in 3196 bursts
in a much shorter (4.3 hours) observation \citep{Z25} of FRB 20240114A 
over which spindown would likely be insufficient to prevent detection of
periodicity.}  Precession of triaxial neutron stars has also been proposed
as a trigger of magnetar activity and an accelerator of the highest energy
cosmic rays \citep{WS24,SW24}.

FRB may be produced by charge bunches moving relativistically along magnetic
field lines directed toward the observer, likely from their neutron stars'
magnetic polar caps \citep{KLB17,BK25}, and beamed like all emission from
relativistically moving charges.  The observed radiation is sensitive to the
orientation of the neutron star and its frozen-in field; detection requires
alignment of the polar cap open magnetic field lines with the line of sight.
By itself, this does not explain the absence of periodicity in FRB, for
radio pulsar emission is also narrowly beamed yet its periodicity is its
most striking feature.
\section{Triaxiality}
{For definiteness, we follow \citet{LL76} in taking the three principal
components of the intertia tensor to be ordered
\begin{equation}
I_3 > I_2 > I_1.
\end{equation}
}
The degree of triaxiality may be parametrized by
\begin{equation}
{\cal T} = {(I_1-I_2)^2(I_1-I_3)^2(I_2-I_3)^2 \over [(I_1-I_0)^2+
(I_2-I_0)^2+(I_3-I_0)^2]^3},
\end{equation}
where the mean moment
\begin{equation}
I_0 = {I_1+I_2+I_3 \over 3}.
\end{equation}
Then
\begin{equation}
0 \le {\cal T} \le {1 \over 2}.
\end{equation}
If the moment tensor is spheroidal (oblate or prolate) two of the $I_i$ are
equal and ${\cal T} = 0$.

${\cal T}$ may approach its maximum value even if the moment tensor is
nearly isotropic (all the $I_i$ are nearly equal), as expected for a slowly
rotating magnetar.  For example, consider
\begin{equation}
\begin{split}
I_1 &= I_0 - \epsilon\\
I_2 &= I_0\\
I_3 &= I_0 + \epsilon.\\
\end{split}
\end{equation}
Then ${\cal T} \to 1/2$ as either $\epsilon \to 0$ or $\epsilon \to I_0$.

{Another parameter measures the deviation of the moment of inertia
ellipsoid from sphericity:
\begin{equation}
{\cal A} = {(I_1-I_0)^2+(I_2-I_0)^2+(I_3-I_0)^2 \over I_0^2}.
\end{equation}
For a triaxial moment tensor
\begin{equation}
0 \le {\cal A} \le 2.
\end{equation}
For the more restricted case of a prolate ellipsoid
\begin{equation}
0 \le {\cal A} \le {3 \over 2}
\end{equation}
while for an oblate ellipsoid
\begin{equation}
0 \le {\cal A} \le {3 \over 8}.
\end{equation}
}

A triaxial moment of inertia tensor may result from frozen-in magnetic
stress $B^2/8\pi = 4 \times 10^{28}B_{15}^2$ dyne/cm$^2$, where the magnetic
field $B = B_{15}\times 10^{15}\,$Gauss.  or frozen-in elastic stress
 ${\cal O}(10^{29}$ dyne/cm$^2$) \citep{HK09}.  These may be of similar
magnitude and each may make the moment of inertia tensor triaxial.
Rotational distortion relaxes to axisymmetry on a fast hydrodynamic or
viscous time scale, and is likely smaller than the other distortions for
rotational periods $\gtrsim 1\,$s.  {This is in contrast to classic
radio pulsars whose magnetic distortion is several orders of magnitude less,i
and that are likely well approximated as oblate spheroids.)

For a magnetically distorted magnetar ${\cal A} \sim 10^{-6}B_{15}^2$.
Frozen elastic anisotropy may provide a comparable value of $\cal A$, but
with the difference that elastic strain relaxes by creep, at unknown rates,
while magnetic fields relax resistively, expected to be very slowly.
The small values of $\cal A$ of magnetars are consistent with ${\cal T} =
{\cal O}(1/2)$, in contrast to radio pulsars whose distortion from
sphericity is rotational and whose dynamical figures are close to oblate
ellipsoids with ${\cal T} \ll 1/2$.}

The orientation of the moment of inertia ellipsoid in an inertial frame
is described by the Euler angles $\theta$, $\phi$ and $\psi$.  These are
related to and vary with the rotation rates $\omega_i$ around the
principal axes $i$ that evolve according to the Euler equations:
\begin{equation}
\begin{split}
{\dot\omega}_1 + {I_3-I_2 \over I_1}\omega_2\omega_3 &= 0\\
{\dot\omega}_2 + {I_1-I_3 \over I_2}\omega_1\omega_3 &= 0\\
{\dot\omega}_3 + {I_2-I_1 \over I_3}\omega_1\omega_2 &= 0.\\
\end{split}
\end{equation}


Solutions of the Euler equations when all moments of inertia are unequal are
complicated and aperiodic \citep{LL76}.  {The Euler angles $\theta$ and
$\psi$, as well as the $\omega_i$, vary periodically (but not sinusoidally)
with a period ${\cal O}(\tau)$ while $\phi$ varies as the sum of two
incommensurate periodic functions whose periods are of that order of
magnitude.  In the words of \citet{LL76} ``the top [rigid triaxial body]
does not at any time return exactly to its original position''.
This, by itself, does not explain the absence of observed periodicities in
repeating FRB because the variations in the $\omega_i$ and Euler angles are
very slow; the implied changes are small during the single (1--4 hours)
continuous observations of repeating FRB in which only upper limits on
periodic modulation have been set.}

To make order of magnitude estimates, replace the time derivatives by the
algebraic factor $1/\tau$, where $\tau$ is a characteristic time scale, the
fractions by $\Delta I/I$, a measure of the asymmetry of the moment of
inertia tensor, and assume $I_1 \approx I_2 \approx I_3$ to high accuracy,
as the case for a dynamically nearly spherical body, but that all the
$|I_i-I_j|$ ($i \ne j$) are of the same order of magnitude.
Then
\begin{equation}
\label{tau}
\tau \sim {I \over \Delta I}{1 \over \omega}.
\end{equation}

The radiation pattern of a hypothetical young triaxial magnetar with $\omega
\sim 10\,$/s (0.6 s period) and $\Delta I/I \sim 10^{-6}$ reorients over
the sky on a time scale $\tau \sim 1\,$day.  {Precession of more slowly
rotating magnetars has been suggested \citep{ZL20,C22,W22a} to explain the
periodic modulations of FRB 20180916B and FRB 20201102A.  Assuming an
intrinsic (but stochastic) mean burst interval $\eta \sim 10^2\,$s, as
observed for active repeaters (that we assume are stably aligned with the
line of sight), this is sufficient to explain the failure to see repetitions
from apparent non-repeaters:  If FRB beams have a width $\sim 0.003\,$rad,
as might be expected if the radiating charges have Lorentz factors $\gamma
\sim 300$, the radiating beam moves out of the line of sight in a few
minutes.  If the beam of an apparent non-repeater has solid angle
$1/\gamma^2 \sim 10^{-5}\,$ sterad and wanders stochastically over all
directions then the implied recurrence time of apparent non-repeaters is
${\cal O}(4 \pi \gamma^2\eta \sim 3\,\text{y})$ and observed duty factors
$\sim 10^{-11}$, consistent with their apparent non-repetition
\citep{U25}.}
\section{Consequences}
\subsection{Repeaters {\it vs.\/} Apparent Non-Repeaters}
I follow \citet{L25,ZH25} in suggesting that in repeating FRB the
magnetic and spin axes are aligned, and \citet{KLB17,BK25} in suggesting
that FRB radiation is emitted along open field lines.  This leads to a
possible explanation \citep{K25a} of the difference between repeating and
apparently non-repeating FRB:  Apparently non-repeating FRB have triaxial
moment tensors and open field lines that are not parallel to their rotation
axis, so that their directions of radiation wander slowly and
pseudo-randomly across much of the sky.

{Because FRB emission is narrowly collimated and FRB
have low emission duty factors (for emission in any direction) $\text{DF} =
\Delta t/\eta \sim 10^{-5}$, where $\Delta t \sim 1\,$ms is a typical burst
width, the chance of observing a second burst (after precession has
misaligned the axis) is small.  FRB that are are only minimally tri-axial
may have their magnetic and rotational axes aligned (dissipation reduces
them to their lowest energy state with oblate spheriodal intertia tensors
and magnetic, emission and rotation axes aligned); if fortuitously
these point to the observer we see them as repeating FRB, while otherwise
they are not observable at all.

The observed duty factor of a triaxial FRB may be ${\cal O}
(\text{DF}\Omega/4\pi) \sim \text{DF}/4\pi\gamma^2$, where $\Omega$ is the
solid angle of emission of the FRB and $\gamma$ the corresponding Lorentz
factor of the emitting charge bunches.  Triaxiality may explain the large
gap between the upper bounds on duty factors of apparently non-repeating FRB
and the measured duty factors DF of repeaters as curvature radiation models
indicate $\gamma^2 \sim 10^5$.

Any apparent non-repeater can be observed, but only during the rare
intervals when its open field lines and emission direction are suitably
aligned, explaining their very small observed duty factors.  In contrast,
only a small fraction of repeaters are aligned so that their bursts can be
observed; the overwhelming majority of repeaters are unobservable.}

If the misalignment is comparable to the beam width then the probability of
detecting a burst depends on how close to the line of sight is the beam axis
because the radiated intensity decreases with increasing deviation.  This
varies aperiodically if the deviations from an isotropic moment tensor are
triaxial (comparable in magnitude, however small).  If the observer remains
close to the beam axis the probability of detecting a burst depends on how
far he is from the axis.  This may explain the absence of periodicity in the
rate of detection of bursts, as well as the absence of an exact (RRAT-like)
underlying period, in active repeating FRB \cite{L21,X21,K22a,K25b}.
\subsection{Varying Activity of Repeaters}
The activity of all repeating FRB varies aperiodically on time scales
from hours to months or years; this is distinct from the 16.35 d periodic
{\it modulation} of activity observed in FRB 20180916B \citep{CHIME} and the
$\approx 160$ d periodic {\it modulation} of activity observed in FRB
20121102A \citep{R20,C21b}.  Periods of intense activity, with bursts
detected at rates of nearly 1000/h \citep{Z25}, are interspersed among
periods in which few or no bursts are detected.   If dissipation has aligned
rotation along the $I_3$ axis intense activity occurs when this axis and
open field lines point to the observer, and low activity when they precess
away.
\subsection{Absence of Burst Periodicity}
If the magnetic and spin axes and the line of sight are closely aligned in
repeating FRB sources \citep{BK25,L25}, rotation about this axis does not
modulate the probability of observing a burst because it does not change the
angle between the magnetic (open field line) axis and the line of sight.
The observed temporal behavior then reflects only the ``plasma weather'' 
(intrinsically stochastic particle acceleration and emission). 
\subsection{Detection of Gravitational Waves From Rotating Neutron Stars}
Eq.~\ref{tau} may also be applied to spinning neutron stars that have been
suggested as sources of detectable gravitational waves.  Candidates include
known fast radio pulsars (Crab and Vela) as well as neutron star X-ray
sources like Sco X-1 hypothesized \citep{G22} to be spun-up by accretion.

In order to emit gravitational waves they must not be axisymmetric.  That
need not, in principle, require them to be triaxial; for example, a
dynamically prolate spheroid spinning about a major axis has a radiating
gravitational quadrupole moment.  However, that geometry is contrived and
implausible.  Such a neutron star is in a high energy state and dissipation
would relax it to a nearly oblate configuration.  If an oblate spheroid, it
would not radiate gravitational waves.

Triaxial deviations (resulting from magnetic or elastic stress) from an
oblate spheroid of a spinning neutron star would be small, likely ${\cal O}
(10^{-6}I_0)$, with ${\cal T} = {\cal O}(10^{-6}GM/(8 \Omega_{rot}^2R^3))
\sim 10^{-3}$ for the Crab pulsar, where $\Omega_{rot}$ is its angular
rotation rate.  Without knowing the initial conditions of
$(\omega_1,\omega_2,\omega_3)$ it is impossible to predict $\omega_1$,
$\omega_2$ and ${\dot \omega}_3$ from Euler's equations, where $\omega_3$
refers to the present rotational axis (assumed to be close to the axis of
greatest moment of inertia), but triaxiality might thwart the coherent
integration required to detect a weak periodic gravitational strain.  This
may be tested for pulsars by coherent integration of the radio signal; if
the radio signal integrates coherently then so will the gravitational
signal. 
\section{FRB {\it vs.\/} Soft Gamma Repeaters}
Although both FRB and Soft Gamma Repeaters (SGR) are associated with
magnetars, SGR emit thermal X-rays and soft gamma-rays from the
photosphere of an equilibrium pair plasma.  This radiates into at least
$2\pi$ sterad ($4\pi$ sterad if the pair plasma fills the magnetosphere).
Anomalous X-ray Pulsar (AXP) emission resembles either SGR emission at much
lower intensity or thermal emission from the neutron star surface.  These
radiate very broadly, but not isotropically, so their emission and
${\cal O}(1)$ rotational modulation is observable from most directions.  It
is insensitive to the neutron star's orientation.  If SGR/AXP are triaxial,
this would not produce behavior like that observed for FRB.
\section{Discussion}
I have argued that three major puzzles about FRB, the orders-of-magnitude
difference between duty factors of repeaters and apparent non-repeaters,
the aperiodic variation in activity of repeaters, and the failure to find
periodic rotational modulation of the burst rate in active repeaters, may be
explained in the neutron stars' moment of inertial tensors are triaxial
(specifically, that all components of the deviation from isotropy are
comparable---the tensor is not close to spheroidal).  If triaxiality
successfully explains these puzzles that would support the model of FRB
emission along open field lines from magnetars' polar caps.

The magnetic fields of the very young (10--100 y old) magnetars suggested
\citep{N25} as sources of FRB may be significantly larger than the
$10^{14}$--$10^{15}$ Gauss observed in Galactic magnetar-SGR inferred to be
$10^3$--$10^4$ years old on the basis of their presence in supernova
remnants of that age.  This would decrease the re-orientation time $\tau$
from the values estimated from the parameters of the older Galactic
magnetars.

Dissipation will tend to align the axis of greatest moment of inertia with
the (conserved) direction of angular momentum.  If effective, this implies
that apparent non-repeaters, with significant triaxiality, are younger than
observed repeaters, whose radiation direction may be nearly constant and
directed to the observer.  Dissipation is offset by excitation by starquakes
and SGR outbursts; it is unclear if dissipation dominates, or on what time
scales.  
\section*{Data Availability}
This theoretical study produced no new data.
\section*{Acknowledgments}
I thank T. Piran for discussions and the Hebrew University, Jerusalem,
Israel for hospitality during this work.

\end{document}